\documentclass[copyright]{eptcs}

\input{header}

\title{Towards the Integration of an \\Intuitionistic First-Order Prover into Coq}
\author{Fabian Kunze
\institute{Saarland University}
\email{s9fakunz@stud.uni-saarland.de}}

\newcommand{\titlerunning}{Towards the Integration of an Intuitionistic First-Order Prover into Coq}
\newcommand{\authorrunning}{Fabian Kunze}

\begin{document}

\maketitle

\begin{abstract}
\noindent An efficient intuitionistic first-order prover integrated into Coq is useful to replay proofs found by external automated theorem provers. We propose a two-phase approach: An intuitionistic prover generates a certificate based on the matrix characterization of intuitionistic first-order logic; the certificate is then translated into a sequent-style proof.
\end{abstract}

\section{Introduction}

Sledgehammer \cite{paulson-blanchette-2010} and HOLyHammer \cite{kaliszyk-urban-2015} drastically improved the productivity for users of proof assistants. They make the capabilities of automated theorem provers (ATPs) available from within interactive proof assistants.

The large, monolithic design of state-of-the-art theorem provers can not be easily trusted to be free of bugs. Thus invoking theorem provers as an oracle is unacceptable for most users. Proof assistants are more trustworthy because all reasoning is checked by a kernel intentionally kept small.

To integrate external provers, small yet efficient, \emph{certified} provers \emph{integrated} into the proof assistant are used: Although it is often possible to mechanically translate the proof to a format accepted by the proof assistant, the integrated prover allows for the reconstruction without the full knowledge of all axioms and rules used by the external prover. Thus an integrated prover simplifies the  integration of not only one but different external provers.

There has been effort to integrate classical provers into Coq, e.g.\ SMTCoq \cite{armand-et-al-2011}, Satallax \cite{brown-2012} and why3 \cite{bobot-et-al-2011}, but they produce proofs that assume classical axioms. As a fair amount of proof developments avoids assuming additional axioms, the acceptance of a future `Coq Hammer' benefits from the integration of an efficient, \emph{intuitionistic} prover.

\section{Existing Intuitionistic Provers in Coq}
\label{exProvers}
The existing intuitionistic first-order provers integrated into Coq are not very strong. We evaluated \texttt{firstorder} \cite{corbineau-2004}, a built-in tactic based on a sequent calculus, and JProver \cite{Schmitt-et-al-2001}, a plugin available for Coq. Using Coq version \texttt{8.6pl1}, we considered first-order problems that are likely to emerge in a future `Coq Hammer'.

For example, we tested formulas where the instantiate of quantifiers is not immediately determined using a goal-driven approach:
\begin{align*}
& \left( \forall x, x = x \right) 
  \land (\forall x, P x \lor Q x) \\
   \land~& (\forall x y, x = y \land P x \Rightarrow R y)
  \land (\forall x y, x = y \land Q x \Rightarrow R y)
  \Rightarrow (\forall x, R x).
\end{align*}

On this formula, \texttt{firstorder} was unable to find a proof during the several minutes we run it. JProver succeeded in less than one second. 

We also invoked both provers on several set-theoretical problems from the ILTP (Intuitionistic Logic Theorem Proving) library \cite{Rath-2007}. Similar to the intended use case, we only supplied the axioms needed for the proofs, resulting in problems like
\begin{align*}
& \left( \forall A B X, X \in A \cup B \Leftrightarrow X \in A \lor X \in B \right) \\
 \land ~&(\forall A B, A = B \Leftrightarrow A \subseteq B \land B \subseteq A) \\
   \land~& (\forall A B, A \subseteq B \Leftrightarrow \forall X, X \in A \Rightarrow X \in B)
  \Rightarrow (\forall A, A \cup A = A).
\end{align*}

On this and similar problems, both \texttt{firstorder} and JProver failed to find proofs before we aborted them after running several minutes.

Therefore, faster intuitionistic provers integrated into Coq are necessary for a `Coq Hammer' used in practice. 
 
\section{Proposed Architecture}

We propose to employ the recent improvements on automated, intuitionistic first-order theorem proving by Otten: ileanCoP \cite{otten-2005,otten-2008} and the forthcoming intuitionistic version of nanoCoP \cite{otten-2016,otten-2016-private}. The existing implementations of both provers verified that the formulas in \autoref{exProvers} are valid in under a second.
Both provers are based on the existence of proof certificates for the matrix characterization of (intuitionistic) validity \cite{Wallen-1990}, which can be translated to sequent-style proofs \cite{Schmitt-1996}.

This architecture is similar to that of JProver (which uses the same characterization of validity), but uses a more efficient proof search procedure, leading to a higher success rate.

\subsection{Finding Proof Certificates}

The performance of ileanCoP is well in identifying valid formulas, compared to other intuitionistic provers \cite{otten-2008}. But it does not keep track of the proof found. Furthermore, it is based on a \emph{clausal} variant of the matrix characterization for intuitionistic logic. The necessary translation into a non-clausal matrix proof has been sketched in the correctness proof of ileanCoP \cite{otten-2005}, but to our knowledge has not yet been implemented.

The classical prover nanoCoP \cite{otten-2016} solves both problems: It outputs the proof certificate found and uses the non-clausal matrix characterization of classical validity. Otten is currently extending nanoCoP to an intuitionistic variant by integrating prefix unification \cite{Wallen-1990}, a method already employed to derive ileanCoP from the classical prover leanCoP.

In our proposed architecture, the proof certificate for a first-order formula $F$ consists of a \emph{multiplicity $\mu$}, a pair of substitutions $\sigma=(\sigma_Q,\sigma_J)$ and a set of pairs of \emph{$\sigma$-complementary} atoms in the formula (connections) that \emph{spans} $F^\mu$.

We will now give an very informal intuition about this certificate.

A Part of the certificate is already needed for the matrix characterization of classical logic:
The multiplicity $\mu$ takes care of the multiple instances an all-quantified subformula of $F$ may be needed in the proof. One part of `$\sigma$-complementary' ensures that that two atoms in a connection are identical under the (non-circular) term substitution $\sigma_Q$, but have different polarity.

The set of connections \emph{spans} the formula if every \emph{path} through the formula contains at least one connections. In the quantifier-free case, each path correspond to a disjunction in the conjunctive normal form. In the case of formulas with quantifiers, each path correspond a branch of a (classical) analytic tableaux, where quantifiers are instantiated according to $\sigma_Q$.

The main difference in the intuitionistic characterization is the use of $\sigma_J$ to ensure that the positions of the pair of complementary atoms in the formula are `compatible'. The position of an atom is defined by structural recursion on the formula and represented by a string, consisting of fresh constants and fresh variables.

An example of this for an intuitionistic valid formula is $P \Rightarrow P$, where the two atoms can be made complementary: The position of the first $P$ is described by the string $z$ with a fresh variable $z$, while the position of the second $P$ is the string $a$ consisting of a fresh constant $a$. Defining $\sigma_J(z)=a$ unifies those strings.

For the formula $\neg P \lor P$, a theorem of classical, but not intuitionistic logic, the two atoms can not be made complementary: The position of the first $P$ is described by $xa$, while the one for the second $P$ is $b$. As the second position contains no variable and no $a$, we can never unify those strings.

This concept generalizes to quantified formulas, but for the main idea, it suffices to study the cases for non-quantified formulas.

For a more formal definition and a few more examples, we recommend the first two Sections of \cite{otten-2005}, and Chapter 8, \S 4 of \cite{Wallen-1990}.

It should be noted that one of the main improvements of nanoCoP compared to JProver is the handling of the multiplicities: nanoCoP adds instances of subformulas during the proof search as needed, while JProver fixes the multiplicity before searching for an proof; on failure, an additional instance of the whole formula gets added and the proof is retried. Although both are complete, the first approach is more goal-driven and thus expected to be more efficient.

\subsection{Generating Sequence Proofs}
The high-level idea is that the proof certificate guarantees that on each branch of the sequent-style proof, eventually complementary atoms are found. The difficulty is to traverse the formulas in the right order, which depends on $\sigma_J$.

The translation of a matrix characterisation proof certificate into a sequent-style proof has already been investigated and implemented for JProver\cite{Schmitt-1996}. We intend to adopt this translation, as we expect it to be reasonable fast: In the examples we tried and where JProver succeeded, the sequence-style proof produced was rather short. In the cases where JProver did not succeed in an acceptable time, it did not even reach the sequence-proof generation. Thus we conclude that the bottleneck of JProver, at least in the examples we tried, is the proof certificate search.

\section{Discussion}

\subsubsection*{Modular vs Monolithic}
We explicitly want to use a modular implementation for the two phases, possibly written in multiple languages. The Prolog version of the intuitionistic variant of nanoCoP is expected to materialize soon and there is already an implementation of the sequence proof generating algorithm integrated into Coq. Thus we expect no challenge in creating a prototype of the suggested architecture using the Prolog program. This would allow us to test whether the proposed setup is suitable for the intended use case.

In the longer term, it would be desirable to have a native OCaml implementation of the proof search procedure, allowing for a deployment within Coq, without additional binaries. The classical leanCoP has been ported to OCaml for the HOL light proof assistant, with performence comparable to the Prolog version\cite{Kaliszyk-2015}. This port can serve as a starting point for a native OCaml version of the forthcoming intuitionistic nanoCoP. Then, the modular approach allows to optionally use external proof procedures. This allows to evaluate improvements to the Prolog proof procedure before porting them.

Also, a modular design allows to more easily use parts an implementation this for other, intuitionistic proof assistants. This additional usage should be kept in mind while developing this, and other, tools towards Hammers in Type Theory.

\subsubsection*{Explicit Proofs vs Reflection}
One approach in proof automation in Coq is `proof by reflection': Some or all parts of the the proof search procedure are written in Coq, including a correctness proof. The proof of a statement then \emph{is} the call to this Coq procedure.

One argument for `proof by reflection' in Coq is the efficiency. But this is just a benefit compared to an implementation using Ltac, the tactic language in Coq: The evaluation of native Coq terms is heavily optimized to the extend of native machine code compilation and execution. In contrast, Ltac is just interpreted on top of several layers of abstraction. As we propose to use OCaml, not Ltac, for the computationally intense parts, this argument does not apply here.

We assume that the search for the proof certificate could be more easily written, modified, or enriched with heuristics, when using a language allowing side effect. This discourages the use of reflection in the first part of our proposed architecture.

Reflection seems to be more reasonable for the second part, the translation to a sequence proof: There is no need to explicitly generate the sequence proof when a certified procedure guarantees that the sequence proof \emph{does} exist when the certificate satisfies the appropriate conditions.

The challenge here would be that the proof certificate must annotated with type information rich enough to reduce to proofs for all formulas we intend to proof: This means that when the terms in the formula are not single sorted, but have of more complex types, e.g. dependent types, this must be incorporated in the proof certificate, the translation procedure itself and its correctness proof. At first, it seems that a benefit would be that the translation is proven to be sound by design. But to check the conditions that a proof certificate is indeed valid is more or less computationally equivalent hard as to generating a sequence-style proof.

Another aspect to consider is that some usage, a formula that is not first-order can be transformed into an first-order formula such that a proof of the later formula can be translated back to a proof of the former formula. In a reflective proof reconstruction, this intermediate steps may can not type-check.

\subsubsection*{Intuitionistic vs Classical}

Automated theorem proving in intuitionistic logic is computationally harder than in classical logic. For developments assuming classical axioms, the intuitionistic part of both phases can be made optional, resembling the classical proof search of nanoCoP without significant overhead.

Note that the proof search in this proposed architecture does neither need skolemization nor clausal normal forms. Thus more structure of the different lemmas and parts of the formulas is preserved and in some sense, this approach is closer to humans reasoning. Further investigation of this architecture could lead to insights useful for automated reasoning in proof assistants of classical logic.
 
 \section*{Acknowledgements}
 We thank Jens Otten for his helpful discussions and suggestions, and Jasmin Blanchette and the anonymous reviewers for their comments on this extended abstract.

\bibliographystyle{eptcs}
\bibliography{literature}


\end{document}